\magnification= \magstep 1
\global\newcount\meqno
\def\eqn#1#2{\xdef#1{(\secsym\the\meqno)}
\global\advance\meqno by1$$#2\eqno#1$$}
%
\global\newcount\refno
\def\ref#1{\xdef#1{[\the\refno]}
\global\advance\refno by1#1}
\global\refno = 1
\vsize=25true cm
\hsize=18 true cm
\hoffset =0.8 true cm
\hsize = 15.8 true cm
\tolerance 10000
%

\def\ln{{\rm ln}}

\def\pr{{\it Phys. Rev.}}
\def\pre{{\it Phys. Rep.}}
\def\re{{\it Rev. Mod. Phys.}}
\def\np{{\it Nucl. Phys.}}
\def\pl{{\it Phys. Lett.}}
\baselineskip 12pt plus 1pt minus 1pt
\centerline{\bf MINI-INSTANTONS IN SU(2) GAUGE THEORY}
\vskip 2cm
\centerline{Vincenzo Branchina\footnote{$^a$}{branchina@crnvax.in2p3.fr} 
and Janos Polonyi\footnote{$^b$}{polonyi@fresnel.u-strasbg.fr}
\footnote{$^c$}{On leave from L. E\"otv\"os University, Budapest, Hungary}}
\vskip 1cm
\centerline{\it Laboratory of Theoretical Physics}
\centerline{\it Department of Physics}
\centerline{\it Louis Pasteur University}
\centerline{\it 67087\ \ Strasbourg\ \ Cedex\ \ France}
\vskip 3cm
\centerline{\bf ABSTRACT}
The effects of instantons close to the cut-off is studied in four dimensional
SU(2) gauge theory with higher order derivative terms in the action. It is 
found in the framework of the dilute instanton gas approximation that
the convergence of the topological observables requires non-universal 
beta function.
\medskip
\vfill
\eject
\xdef\secsym{}\global\meqno = 1
Only renormalizable Quantum Field Theory models are considered in 
Particle Physics. This was explained traditionally by 
inspecting the UV divergences generated by 
the operators in the framework of the perturbation expansion 
on a homogeneous background field. The non-renormalizable
theories were rejected due to the need of infinitely many coupling constants. 
This argument has been further developed in the last 
decades. First it came the realization that what really matters 
in Particle Physics is not the true UV divergence because one 
always works with effective theories in the lacking of definite knowledge 
of the Theory of Everything. The characterization of the operators 
according to the renormalizability was replaced by the procedure of 
identifying their importance at
low energies. In particular, the equivalence of the renormalizability
of an operator with its relevance at the UV fixed point of the theory
has been established \ref\wilsrg. The non-renormalizable operators were
excluded in this manner because they do not change the universality class, i.e. their
influence on the dynamics decreases as we move
away from the UV fixed point towards the physical energy scales. 
An important feature of this new characterization of the class of 
important operators is that it is
based on one scaling regime, that of the UV fixed point. It may happen that
there are other scaling regimes as we move towards the IR and certain
non-renormalizable operators become relevant there \ref\senben. 
One should include these operators in the description 
of the physics at finite energies. 
The second improvement in defining the class of acceptable theories of Particle 
Physics came by extending the computation of the scaling
laws of the UV fixed point beyond the realm of the perturbation expansion.
The usual strategy is to take the anomalous dimensions into account
in the power counting argument \ref\bard. New relevant operators of the UV 
fixed point can be found in this manner and the class of parameters which
characterize the physics of a given particle content is enlarged.

The physical picture of the strong coupling massless QED vacuum 
\ref\miransky\ which served as the motivation of Ref. \bard\ is based on the 
observation that the positronium may acquire
negative energy and collapse onto the size of the cut-off when $e=O(1)$. 
The condensate of these bound states breaks the chiral symmetry and the
resulting vacuum is modified in the IR domain when 
compared to the perturbative one.
A slight modification of this scenario, namely the generation of a phase 
transition by localized coherent states at the cut-off has been suggested in 
Refs. \ref\enzo, \ref\jochen.

The aim of this Letter is to show that these coherent states,
namely the solutions of the equation of motion in the presence of higer 
derivative terms, modify the physics in a manner which is not predicted 
by the usual power counting based on the expansion around a locally flat
background field. The important point is that this is a 
{\it tree level} effect which is due to the sensitivity of these solutions 
to the presence of the higer order derivative terms in the action. 
The usual power counting method and its more sophisticated version the,
decoupling theorem \ref\apcar, are constructed from the very beginning 
to trace the effects of the quantum fluctuations at the loop level and can 
not detect such an effect. We show that the action of the localized coherent 
states can be lowered such that the long range physics and the beta function 
are dominated by mini-instantons, i.e. instantons at the cut-off scale. 

In our previous paper \enzo\ the non linear sigma model was studied and a
non-universal dependence of the beta function was found. But this result
might have aroused from the fact that even the usual, large 
instanton gas is UV divergent in this model. 	
The study of the four dimensional SU(2) gauge model where the usual 
instanton gas is UV finite demonstrates this genuinely new, cut-off
independent tree level effect and brings us closer to an eventual 
phenomenological application of this phenomenon.

It is rather straightforward to
show that the mini-instantons modify the scaling laws for topological
observables by means of the dilute gas approximation. What remains an
open question is to
find out if the interactions between the instantons generate the same
scaling laws for topological and non-topological observables.

In order to take into account the contributions of the saddle points 
in the vicinity of the cut-off one has to turn to the bare theory.
So we start with the comparison of the renormalized and the bare perturbation 
expansions. The latter refers to a well defined large but finite dimensional
path integral of the bare theory which is defined by the bare action,
$S_B=S_{B0}+S_{B1}$,
\eqn\bpe{Z=\int D[A]e^{-S_{B0}[A]-S_{B1}[A]}
=\sum_n{(-1)^n\over n!}\int D[A]e^{-S_{B0}[A]}(S_{B1}[A])^n.}
The convergence of the expansion in the bare coupling constant, $g^2_B$, 
usually turns out to be asymptotic only. Another problem of the expansion
is that each order diverges with the cut-off and can not really be considered
as small. The renormalized perturbation expansion is based on the 
observation that for the suitable fine tuned bare parameters the divergences cancel
order by order and the resulting sum can be reorganized by means of a new,
renormalized small parameter, $g^2_R(\mu)$, which is independent of the cut-off. 
To achieve this 
one splits the bare action into the renormalized and the counterterm parts,
$S_B=S_{R0}+S_{R1}+S_{CT}$ and writes
\eqn\rpe{Z=\sum_n{(-1)^n\over n!}
\int D[A]e^{-S_{R0}[A]}(S_{R1}[A]+S_{CT}[A])^n.}
But note that $g^2_R(\mu)$ is constructed only $after$ using the Wick theorem
to find the cancellation between $S_{R1}$ and $S_{CT}$. The expansion is formal 
because there is no well defined path integral with $g^2_R(\mu)$ as small 
parameter.
The smallness of $g^2_R(\mu)$ is misleading because it hides the problem that its very 
existence requires the cancellation of divergences $after$ expanding the bare integrand.
The question of the convergence or the applicability of the expansion can only be 
investigated on the bare level.

The saddle point expansion is similar to the straightforward perturbation expansion
from this point of view. The bare expansion starts with 
\eqn\bse{Z=\sum_{A^{(B)}_{cl}}e^{-S_B[A^{(B)}_{cl}]}
\int D[A]e^{-{1\over2}A({\delta^2\over
\delta A\delta A}S_B[A^{(B)}_{cl}])A+O(A^3)},}
where ${\delta S_B[A^{(B)}_{cl}]\over\delta A}=0$
and continues with the expansion in the $O(A^3)$ pieces of the action. On the 
contrary, the renormalized saddle point expansion \ref\thooft\ uses $S_R$
to select the saddle points and the rest of the action, $S_{CT}$, appears 
on the loop corrections only,
\eqn\rse{Z=\sum_{A^{(R)}_{cl}}e^{-S_R[A^{(R)}_{cl}]}
\int D[A]e^{-{1\over2}A({\delta^2\over\delta A\delta A}(S_R[A^{(R)}_{cl}]
+S_{CT}[A^{(R)}_{cl}]))A+O(A^3)},}
with the choice ${\delta S_R[A^{(R)}_{cl}]\over\delta A}=0$. 
The one-loop integral which arises by retaining the Gaussian part of the
exponent of the integrand is divergent. If the saddle points are
large, i.e. their size parameter is independent of the cut-off then the counterterms
needed to remove the divergences are the same as in the straightforward
perturbation expansion because the instantons are locally flat. 

But note that the configuration $A^{(R)}_{cl}$ is {\it not}
an extremum of the complete integrand and \rse\ is incomplete, the correct 
expression which takes into account $S_{CT}[A]$ reads as
\eqn\rsec{\eqalign{Z&=\sum_{A^{(R)}_{cl}}e^{-S_R[A^{(R)}_{cl}]
-S_{CT}[A^{(R)}_{cl}]}\cr
&\times\int D[A]e^{-A{\delta S_{CT}[A^{(R)}_{cl}]\over\delta A}
-{1\over2}A({\delta^2\over\delta A\delta A}(S_R[A^{(R)}_{cl}]
+S_{CT}[A^{(R)}_{cl}]))A+O(A^3)}.}}
The difference between \rse\ and \rsec\ is that the tree level scale dependence
of $S_{CT}[A]$ is retained in the latter. Is this really important ? One 
would think that for an asymptotically free theory like the Yang-Mills system 
$S_R[A]>>S_{CT}[A]$ 
and the tree level scale dependence is not important. In fact, one has
\eqn\msr{S_R[A]=-{1\over4g^2_R(\mu)}\int d^4x(F^a_{\mu\nu})^2,}
\eqn\msc{S_{CT}[A]=-{\beta_0\over2}\ln{\Lambda\over\mu}
\int d^4x(F^a_{\mu\nu})^2,}
with $\beta_0={22\over3}{1\over16\pi^2}$ in dimensional regularization 
what yields
\eqn\csall{{1\over g^2_B}={1\over g^2_R(\Lambda)}={1\over g^2(\mu)}
+\beta_0\ln{\Lambda^2\over\mu^2},}
and consequently
\eqn\egycs{{1\over g^2(\mu)}>>\beta_0\ln{\Lambda^2\over\mu^2}.}

The problem with this argument is that \msc\ is incorrect. The 
scale dependence which is given by the logarithmic expression in \msc\ comes
from the one-loop level. The analytical regularization provides the cut-off for the
loop integration without having it at the tree level where the actual path 
integral is defined. It is obvious that $S_{CT}[A]$ must be scale dependent since 
it is an irrelevant operator with the role to suppress the high energy modes. 
The scale factor $\mu$ in \msc\ comes from the loop integration and 
is actually the characteristic scale of the configuration $A$ and can not be 
considered as an independent parameter in \msc. Due to the lack of the tree
level scale dependence in \msc\ \egycs\ does not imply $S_R[A]>>S_{CT}[A]$.
The summary: The dimensional regularization
is inconsistent in the presence of a nonhomogeneous background field
because it provides the scale dependence only for the quantum 
fluctuations and not for the tree level action. A mathematically well defined 
regulator which yields a large but finite dimensional integral, \bse, must 
regulate the whole path integral, both at the tree and the loop orders. 

This problem is obvious in lattice regularization when $S_R[A]$ is chosen
to be the classical action in the IR regime, i.e. classical continuum action. 
The lattice and the continuum action agree for slowly varying field 
configurations but differ significantly 
when the characteristic length scale of the configuration is 
close to the lattice spacing. It is just this tree level scale dependence what 
serves as the regulator.

There is another reason why the tree level contribution of the counterterms 
is important in four dimensional Yang-Mills theories. Since the renormalized
action is scale invariant its saddle points, $A^{(R)}_{cl}$, are degenerate when
the tree level scale dependence of $S_{CT}[A]$ is ignored. But one should bear 
in mind that an 
infinitesimally weak splitting of an infinitely degenerate situation can be
important because it may induce non-analytic behavior as it is well
know from the quantum Hall effect.

We know now other consistent regularization method than putting the theory
on the lattice. The difficulties of the analytical approach in lattice 
regularization forces one to find a compromise, namely a variant of the
renormalized action where the counterterms have less role to play. In
such a theory the renormalized expansion is more reliable.
In particular, we suggest the use of higher power of the derivatives in the
renormalized lagrangian. The new pieces will render the higher loop 
contributions finite so the counterterm will be needed for the one-loop
graphs only.

We shall consider two versions of SU(2) gauge theories which differ
in regularization only. One is given in the continuum by the action
with higher order derivatives,
\eqn\yml{L_{ext}=-{1\over4g_B^2}F^a_{\mu\nu}\biggl(1+{c_2\over\Lambda^2}D^2
+{c_4\over\Lambda^4}(D^2)^2\biggr)F^{a\mu\nu}}
where $c_\alpha>0$ and $D$ is the covariant derivative. 
\yml\ is not the complete bare
lagrangian as it stands because the higher order derivative terms do not remove 
the one-loop divergences. One can use Pauli-Villars regularization 
\ref\slavnov, $L_B=L_{ext}+L_{PV}$,
where $L_{PV}$ stands for the lagrangian of the regulator particle
of mass $M^2=O(\Lambda^2)$ to render the one-loop
structure finite. The contribution of states with negative norm
can not be treated consistently in the path integral
because the Gaussian integral diverges exponentially. The usual
remedy of this problem is to consider the analytical continuation 
of the free generator functional and to change the sign of the propagator
for the regulator particle. The tree level contribution of the regulator is 
lost in this manner and only the loop contributions are retained from $L_{PV}$. 
Our point is that choosing $L_R=L_{ext}$ and $L_{CT}=L_{PV}$
in the renormalized saddle point expansion yields better results
with $c_\alpha\not=0$ than with $c_\alpha=0$. This is because
all higher loop contributions are finite for $L_{ext}$
and the non-consistency comes from the terms which influence the
tree and one-loop level only.
 
Another regularization we shall consider is to put the theory on a space-time 
lattice with $\Lambda={2\pi\over a}$. The
higher order derivative terms give rise Wilson loops up to the length
of $3a$ in the lattice action $L_B=L_{lat}$. The theory is fully 
regulated but not well suited to the analytical methods.

The theory \yml\ possesses saddle points which become self dual
instantons for $c_\alpha=0$. According to the dimensional argument
the tree level instanton action must be of the form
\eqn\confa{{1\over g^2}S_{ext}(\rho\Lambda)={1\over g^2}S_0
\biggl(1-{\tilde c_2\over\rho^2\Lambda^2}
+{\tilde c_4\over\rho^4\Lambda^4}\biggr)}
where $S_0=8\pi^2$ is the scale independent usual instanton action.
$\rho$ is the scale parameter of the instanton configuration 
$A_\mu^{(c_\alpha,\rho)}(x)$ and $\tilde c_\alpha=O(c_\alpha)$. 
The scale parameter $\rho$ is introduced in a somehow arbitrary fashion
by requiring that $A_\mu^{(c_\alpha,\rho)}(x)$ be a self dual instanton with
size $\rho$ for $c_\alpha=0$. But it is only the actual relation between 
$c_\alpha$ and $\tilde c_\alpha$ which depends on the details of introducing 
the scale parameter. We change the notation slightly now and the $1\over g^2$ 
will be factorized out from the action in the rest of the paper.
The stable saddle point corresponds
to an instanton at the scale of the cut-off,
\eqn\insts{\bar\rho={1\over\Lambda}\sqrt{2\tilde c_4\over\tilde c_2},}
with action
\eqn\insta{\bar S_{ext}=S_0\biggl(1-{\tilde c_2^2\over4\tilde c_4}\biggr).}
Since $S_{ext}(\rho\Lambda)$ has vanishing curvature at its minimum
there is no restoring force for the small fluctuations in $\rho$.
Consequently $\rho$ will be used to parametrize a one parameter
family of configurations to be treated by the collective coordinate 
method in the path integration even when the tree level scale invariance 
is removed by $c_\alpha\not=0$.

The scale parameter can be introduced in lattice regularization, too. 
By the help of some well defined but not unique procedure on can
define interpolating field in the continuum which gives the desired link
variables. The scale parameter is then introduced for the interpolating
field. The cut-off dependence in the fully regulated lattice action prevents
us from obtaining a simple expression for the instanton action 
$S_{lat}(\rho\Lambda)$. The tree level matching of the two regularizations
gives 
\eqn\trlm{S_{lat}(\rho\Lambda)=S_{ext}(\rho\Lambda)
\biggl(1+O\bigl((\rho\Lambda)^{-2}\bigr)\biggr).}
We are interested in lattice theories which are similar to the
case of $c_\alpha>0$ i.e. where $S_{lat}(\rho\Lambda)$ reaches the minimum
value $\bar S_{lat}$ at $\bar\rho=O(\Lambda^{-1})$.

Let us compute the ratio of the partition function in the unit and the
zero winding number sector in the one-loop approximation,
\eqn\ratg{{Z_1\over Z_0}={C\over g^{n_0}}\int d^4R
\int_0^{1\over\mu}{d\rho\over\rho^5}
e^{-{1\over g^2}S(\rho\Lambda)}{\cal D}(\rho\Lambda),}
where $n_0$ is the number of modes treated by the help of the collective
coordinate method and ${\cal D}$ stands for the contribution of the 
fluctuation determinant. $C$ is a numerical constant whose value is
of no importance for us.
The infrared catastrophe is ignored by the introduction of the infrared
cut-off, $\mu=O(\Lambda_{QCD})$. Four inverse power of the scale
parameter in the integrand is to give the correct entropy, i.e. number of 
time an instanton with size $\rho$ can be placed into the quantization volume
$V=\int d^4R$. We shall compute \ratg\ for the Pauli-Villars and the lattice 
regulated theory.

In the Pauli-Villars regulated continuum theory the translation in
space-time is a zero mode together with the global gauge transformations
so
\eqn\ratiopv{\biggl({Z_1\over Z_0}\biggr)_{PV}=C{V\over g^8}\int_0^{1\over\mu}
{d\rho\over\rho^5}e^{-{S_0\over g^2}\bigl(1-{\tilde c_2\over\rho^2\Lambda^2}
+{\tilde c_4\over\rho^4\Lambda^4}\bigr)}{\cal D}_{PV}(\rho\Lambda).}

The collective coordinates, $R^\mu$ and $\rho$, are introduced in lattice 
regularization to monitor the continuous space-time translations and the scale 
transformation of the interpolating field configuration, respectively. 
The continuous translational symmetry is broken in the lattice regulated theory. 
The instanton action, $S_{lat}(\rho\Lambda,R^\mu\Lambda)$, and the fluctuation 
determinant, ${\cal D}_{lat}(\rho\Lambda,R^\mu\Lambda)$, become nontrivial 
functions of the translation with period length ${2\pi\over\Lambda}$. 
Since it is sufficient to make the integration over the translations within the
hypercube of the size of the period length one finds for the lattice 
volume $N^4$,
\eqn\ratiol{\eqalign{\biggl({Z_1\over Z_0}\biggr)_{lat}
&=C{N^4\over g^8}\int d^4R\int_0^{1\over\mu}{d\rho\over\rho^5}
e^{-{1\over g^2}S_{lat}(\rho\Lambda,R^\mu\Lambda)}
{\cal D}_{lat}(\rho\Lambda,R^\mu\Lambda)\cr
&=C{V\over g^8}\int_0^{1\over\mu}{d\rho\over\rho^5}
e^{-{1\over g^2}S_{lat}(\rho\Lambda)}{\cal D}_{lat}(\rho\Lambda),}}
where the space time averaging is made implicit by the introduction of  
$S_{lat}(\rho\Lambda)$ and ${\cal D}_{lat}(\rho\Lambda)$,
\eqn\spav{\biggl({2\pi\over\Lambda}\biggr)^4
e^{-{1\over g^2}S_{lat}(\rho\Lambda)}
=\int d^4Re^{-{1\over g^2}S_{lat}(\rho\Lambda,R^\mu\Lambda)}.}
and
\eqn\spavd{\biggl({2\pi\over\Lambda}\biggr)^4
e^{-{1\over g^2}S_{lat}(\rho\Lambda)}{\cal D}_{lat}(\rho\Lambda)
=\int d^4Re^{-{1\over g^2}S_{lat}(\rho\Lambda,R^\mu\Lambda)}
{\cal D}_{lat}(\rho\Lambda,R^\mu\Lambda).}

The contribution of instantons as the function of the size parameter
has a well pronounced peak at $\rho\approx\bar\rho$. Another important region
is in the infrared where the loop correction increase.
In order to separate the contribution of the large, i.e. cut-off
independent instantons from that of the stable saddle point in the
vicinity of the cut-off we split the scale integration into two parts
by the help of a scale parameter $m$, $\Lambda_{QCD}<<m<<\Lambda$,
\eqn\intspl{\int_0^{1\over\mu}d\rho\cdots=\int_0^{1\over m}d\rho\cdots+
\int_{1\over m}^{1\over\mu}d\rho\cdots.}
The first and the second integral will be referred as the contribution 
of the mini and the large instantons, respectively.

{\it Large instantons:} 
\eqn\large{\biggl({Z_1\over Z_0}\biggr)_{PV,large}=
C{V\over g^8}\int_{1\over m}^{1\over\mu}{d\rho\over\rho^5}e^{-{S_0\over g^2}
\bigl(1-{\tilde c_2\over\rho^2\Lambda^2}+
{\tilde c_4\over\rho^4\Lambda^4}\bigr)}{\cal D}_{PV}(\rho\Lambda).}
In this case ~$\rho^2\Lambda^2~>> 1$~ and we can neglect the terms multiplying
$\tilde c_\alpha$ in the exponential. Moreover the quantum fluctuation 
determinant is ${\cal D}_{PV}(\rho\Lambda)=c_{PV}(\rho\Lambda)^{22\over3}$
in this regime. We finally obtain
\eqn\larg{\eqalign{\biggl({Z_1\over Z_0}\biggr)_{PV,large}
&=C_{PV}{V\over g^8}\int_{1\over m}^{1\over\mu}{d\rho\over\rho^5} 
e^{-{S_0\over g^2}}(\rho\Lambda)^{22\over3}\cr
&={3\over10}C_{PV}{V\Lambda^4\over g^8}e^{-{S_0\over g^2}}
\Biggl[({\Lambda\over\mu})^{10\over3}-({\Lambda\over m})^{10\over3}\Biggr]
\cr
&\sim{3\over10}C_{PV}{V\Lambda^4\over g^8}
e^{-{S_0\over g^2}}({\Lambda\over\mu})^{10\over3},\cr}}
where $C_{PV}=Cc_{PV}$.

In lattice regularization we use \trlm\ and
\eqn\detm{{\cal D}_{lat}(\rho\Lambda)={\cal D}_{PV}(\rho\Lambda)
\biggl(c_{lat}+O\bigl((\rho\Lambda)^{-2}\bigr)\biggr),}
to arrive at
\eqn\largl{\biggl({Z_1\over Z_0}\biggr)_{lat,large}
\sim{3\over10}C_{lat}{V\Lambda^4\over g^8}
e^{-{S_0\over g^2}}\biggl({\Lambda\over\mu}\biggr)^{10\over3},}
with $C_{lat}=C_{PV}c_{lat}$.

{\it Mini-instantons:}
\eqn\smallpv{\eqalign{\biggl({Z_1\over Z_0}\biggr)_{PV,mini}
&=C{V\over g^8}\int_0^{1\over m}
{d\rho\over\rho^5}e^{-{S_0\over g^2}
\bigl(1-{\tilde c_2\over\rho^2\Lambda^2}+
{\tilde c_4\over\rho^4\Lambda^4}\bigr)}{\cal D}_{PV}(\rho\Lambda)\cr
&=C{V\over g^8}e^{-{S_0\over g^2}
\bigl(1-{\tilde c_2^2\over4\tilde c_4}\bigr)}
\int_0^{1\over m}{d\rho\over\rho^5}
e^{-{S_0\over g^2}\tilde c_4\bigl({1\over\rho^2\Lambda^2}-
{1\over\bar\rho^2\Lambda^2}\bigr)^2}{\cal D}_{PV}(\rho\Lambda)\cr
&=C{V\over g^8}e^{-{1\over g^2}\bar S_{ext}}\int_{m^2}^\infty
d\Bigl({1\over\rho^2}\Bigr){1\over\rho^2} {\cal D}_{PV}(\rho\Lambda)
e^{-{S_0\over g^2}\tilde c_4
\bigl({1\over\rho^2\Lambda^2}-{1\over\bar\rho^2\Lambda^2}\bigr)^2}\cr
&\sim C{V\Lambda^4\over g^8}e^{-{1\over g^2}\bar S_{ext}}
{{\cal D}_{PV}(\bar\rho\Lambda)\over\bar\rho^2\Lambda^2}
\sqrt{\pi g^2\over S_0\tilde c_4}\cr
&=C{V\Lambda^4\over g^8}e^{-{1\over g^2}\bar S_{ext}}
{\cal D}_{PV}(\bar\rho\Lambda)
\sqrt{{\tilde c_2^2\over4\tilde c^3_4}{g^2\over 8\pi}}}}
Similar steps followed in the lattice regularized model yield
\eqn\smallpv{\eqalign{\biggl({Z_1\over Z_0}\biggr)_{lat,mini}
\sim C{V\Lambda^4\over g^8}e^{-{1\over g^2}\bar S_{lat}}
{\cal D}_{lat}(\bar\rho\Lambda)\biggl({1\over g^2}
{d^2S_{lat}(\bar\rho\Lambda)\over d(1/(\rho\Lambda)^2)^2}\biggr)^{-1/2}.}}
Note that both the large and mini instanton contributions are well
defined since they are independent of the choice of $m$.

Finally we take the limit $\Lambda\to\infty$
by keeping a renormalization condition fulfilled for each relevant 
coupling constant. Suppose that the usual scenario holds, namely that $F^2$ 
is the only relevant operator, \csall\ is valid,
${S_0\over g^2(\Lambda)}={22\over3}\ln({\Lambda\over\Lambda_{PV}})$,
and $c_\alpha$ are irrelevant so need no renormalization. The renormalization
condition to fix $g^2$ will be chosen as
\eqn\rcz{\Lambda{d\over d\Lambda}{Z_1\over Z_0}=0.}

In the limit $\Lambda\to\infty$ either the large or the mini instantons
dominate the partition function. In order to see this we compute
their relative weight,
\eqn\ratio{\eqalign{R_{PV}&={\biggl({Z_1\over Z_0}\biggr)_{PV,large}\over
\biggl({Z_1\over Z_0}\biggr)_{PV,mini}}=
{3\over10}{e^{-{S_0\over g^2}}
\biggl({\Lambda\over\mu}\biggr)^{10\over3}\over
e^{-{S_0\over g^2}\bigl(1-{\tilde c_2^2\over4\tilde c_4}\bigr)}
{\cal D}_{PV}(\bar\rho\Lambda)
\sqrt{{\tilde c_2^2\over4\tilde c^3_4}{g^2\over 8\pi}}}\cr
&={\rm Const}\times\ln~{\Lambda\over\Lambda_{PV}}
\biggl({\Lambda_{PV}\over\mu}\biggr)^{{22\over3}(1-{\bar S_{ext}\over S_0})}
\biggl({\Lambda\over\mu}\biggr)^{{2\over3}(11{\bar S_{ext}\over S_0}-6)}.}}
For $\bar S_{ext}<S_0{6\over11}$ the mini-instantons dominate the partition 
function.

For the lattice regulated theory where 
${S_0\over g^2}={22\over3}\ln({\Lambda\over\Lambda_{lat}})$ we find
\eqn\ratiol{R_{lat}={\rm Const}\times\ln~{\Lambda\over\Lambda_{lat}}
\biggl({\Lambda_{lat}\over\mu}\biggr)^{{22\over3}(1-{\bar S_{lat}\over S_0})}
\biggl({\Lambda\over\mu}\biggr)^{{2\over3}(11{\bar S_{lat}\over S_0}-6)}.}
The mini-instantons dominate if $\bar S_{lat}<S_0{6\over11}$.

Our renormalization condition, \rcz, can be written in the large instanton
dominated region in the leading order by ignoring the power dependence
in $g^2$ as
\eqn\rgzl{\Lambda{d\over d\Lambda}\biggl(\Lambda^{22\over3}
e^{-{1\over g^2}S_0}\biggr)=\biggl({22\over3}+{2S_0\over g^3}\beta(g)\biggr)
\Lambda^{22\over3}e^{-{1\over g^2}S_0}=0.}
Thus we find $\Lambda{d\over d\Lambda}g=\beta(g)=-{11\over3S_0}g^3$,
which is the first universal term of the usual beta function. This is
by no means surprising, since the ultraviolet structure is the same
on the flat or the large instanton background. 

In the case of the mini-instantons the fluctuation determinant is cut-off
independent and the remaining pieces of the renormalization condition yield
\eqn\rgzm{\Lambda{d\over d\Lambda}\biggl(\Lambda^4
e^{-{1\over g^2}\bar S}\biggr)=\biggl(4+{2\bar S\over g^3}\beta(g)\biggr)
\Lambda^4e^{-{1\over g^2}\bar S}=0,}
in either regularization. Thus we arrive at a different result,
$\Lambda{d\over d\Lambda}g=-{2\over\bar S}g^3$.

What we found is that either $c_\alpha$ are relevant
and need renormalization or the beta function for $g$ is non-universal
in the mini-instanton dominated phase of the theory. This conclusion
depends crucially on the renormalization condition, \rcz. This equation seems
reasonable in the large instanton case but one may object that 
$Z_1/Z_0$ is unimportant in the mini-instanton phase. In fact, $Z_1$ is 
saturated by a mini-instanton and these configurations should not be important 
for observables such as Wilson loops at finite length scale. But this argument 
to exclude mini-instantons from the renormalization process is too superficial.

One can distinguish two different classes of observables in gauge theories. 
The first class contains the quantities with non-topological origin, such as the 
Wilson loops. The second class consists of observables which 
depend on the topological properties of the configuration, such as the 
topological charge, mass of the $\eta'$ meson or the 
Green functions for massless fermions. These quantities are discontinous,
locally constant step functional of the gauge field. 
The mini-instantons play no direct
role in determining the value of an observable of the first class when this
latter it is evaluated on a single gauge field configuration. This is because 
the configuration becomes pure gauge too fast as we move away from the 
instanton and gives negligible contribution to large Wilson loops. 
But the ratio between the instanton size and the length
scale of the observable is irrelevant in case of the topological quantities
and mini-instantons influence the observables of the second class. In fact, the 
topological properties such as as winding numbers can be read off from the 
boundary conditions in space-time, infinitely far from the source of the 
nontrivial topological structure. For example the chiral condensate measured 
in the infrared limit contains the contribution of instantons at all length 
scale. The dilute instanton gas picture suggests that this feature, namely the 
unimportance or importance of mini-instantons for observables in the 
first or second class, respectively, remains valid after averaging 
over the configurations. If this is true then we can not remove the cut-off
by keeping both classes convergent in the mini-instanton phase because 
they require different beta functions. 

But this is certainly an oversimplification suggested by the dilute 
instanton gas picture where the topological structure and the
dynamics are largely unrelated. The interaction between mini-instantons may 
modify the long range structure in a manner what is similar to the effect of 
polarization in classical electrodynamics of continuous media. There are
two possibilities: (i) The topological properties are lost during the
renormalization and the beta function of the theory is given by the
non-topological observables. (ii) The topology is preserved as the
cut-off is removed and the instanton-instanton interactions generate
a uniform beta function what keeps both kind of observables finite. Thus the 
actual renormalization of the theory requires a more detailed computation 
than the dilute gas approximation in the mini-instanton phase. 
It is sufficient to mention here the result of Ref. \jochen\
where the SU(2) lattice gauge model with higher order derivatives was studied 
and it has been found that the vacuum becomes a crystal of frustrations when 
the action may take values below those of the trivial vacuum. This crystal
serves as an example to demonstrate that the strong correlations between the 
localized saddle points may generate new phase diagram and scaling law.

Our simple classification of the observables presented above is
straightforward in the continuum where the topological properties of the
space-time are explicit. But the difference between the topological sectors is
washed away in lattice regularization because the consistent regularization
removes the singularities even on the tree level. The mini-instantons
display topological effects on the lattice only if they are
larger than the lattice spacing, $\bar\rho>\ell_0/\Lambda$. The important point 
is that this condition can be satisfied by a cut-off independent choice of
$\bar\rho\Lambda$. In other words, for the appropriate choice of $c_\alpha$
the mini-instanton size is few order or magnitude larger than the lattice 
spacing but remains proportional with it. The scaling properties of such a 
theory are non-universal. 

One can not keep the topological and the non-topological
observables fixed during the renormalization in the 
mini-instanton phase unless the correlation between the instantons
is taken into account. Thus all one can say at this time is that the
universality for topological quantities like the $\eta'$ mass is not yet 
established in QCD. We may take lattice QCD with Wilson action as an
example where there is no stable instanton solutions but the small
instanton action is so low that the topological susceptibility diverges. By the 
help of adding larger Wilson loops to the action we may generate sufficiently 
large mini-instantons and the universality becomes questionable.
\bigskip
\centerline{\bf REFERENCES}
\medskip
\nobreak
\item{\wilsrg}K. Wilson and J. Kogut, \pre {\bf 12 C} (1974) 75 ;
K. Wilson, \re {\bf 47} (1975) 773. 
\item{\senben} S. B. Liao, J. Polonyi, \pr {\bf D51} (1955) 4474.
\item{\bard}C.N. Leung, S.T. Love and William A. Bardeen,
\np {\bf B 323} (1989) 493 ;
C.N. Leung, S.T. Love and William A. Bardeen, \np {\bf B273} (1986) 649.
\item{\miransky}P.I. Fomin, V.P. Gusynin and V.A. Miranski, \pl {\bf 78B}
(1978) 136; V. A. Miransky, {\it Dynamical Symmetry Breaking in Quantum Field 
Theories}, Singapore, World Scientific, 1994.
\item{\enzo} V. Branchina, J. Polonyi, \np (1955) 99.
\item{\jochen} J. Fingberg and J. Polonyi, {\it "Anti-Ferromagnetic
Condensate in Yang-Mills Theory"}, hep-lat/9602003, submitted to \np.
\item{\apcar} T. Appelquist, J. Carazzone, \pr {\bf D11} (1975) 2856.
\item{\thooft}G. 't Hooft, \pr {\bf D 14} (1976) 3432;
A. Polyakov, \np {\bf B 120} (1977) 429.
\item{\slavnov} T. D. Bakeev and A. A. Slavnov, {\it "Higher Covariant
Derivative Regularization Revisited"}, hep-th/9601092.
\vfill
\eject
\end